\documentstyle[12pt]{article}
\newcommand{\beq}{\begin{equation}}
\newcommand{\eeq}{\end{equation}}
\newcommand{\BE}{\begin{equation}}
\newcommand{\BEAL}{\begin{eqnarray}}
\newcommand{\EE}{\end{equation}}
\newcommand{\EEAL}{\end{eqnarray}}
\def\apj{{\it Astrophys. J.}}
\def\apjl{{\it Astrophys. J. Lett.}}
\def\_#1{_{\scriptscriptstyle #1}}
\def\^#1{^{\scriptscriptstyle #1}}
\def\ecmt{\rm{erg}\,\rm{cm}^{-2}}
\def\gr{$\gamma$-ray}
\def\g{$\gamma$}
\def\es{\,\rm{erg}\,\rm{sec}^{-1}}
\def\ecms{\,\rm{erg}\,\rm{cm}^{-2}\,\rm{s}^{-1}}

\def\ev{\,\rm{eV}}

\def\cc#1{$^{\cite{#1}}$}

\def\crd{cosmic-ray}
\def\crs{cosmic rays}
\def\grb{\gr\ burst}
\def\grbd{\gr-burst}
\def\grbs{\gr\ bursts}

\begin{document}
\begin{center}
\vskip 0.5truein
{\bf  GAMMA-RAY BURSTERS AS SOURCES OF COSMIC RAYS}\\
\vskip 0.5truein
Mordehai Milgrom and Vladimir Usov \\
 Department of condensed-matter physics,
Weizmann Institute, Rehovot 76100 Israel     \\
\end{center}
\vskip 0.5truein
\vskip 30pt
\baselineskip 12pt
%\begin{abstract}
%\end{abstract}
%\keywords{Gamma-ray bursts, cosmic rays}

%\section{Introduction}
\par
{\bf The little we do know of the physical conditions in \gr\
 bursters\cc{har}
 makes them conducive to the acceleration of high-energy cosmic
 rays (CRs), especially if they are at cosmological distances.
 We find that, with the observed statistics and fluxes of \grbs,
 cosmological bursters  may be
 an important source of \crs\ in two regions of the observed spectrum:
 1. At the very-high-energy end ($E>10^{19}\ev$), where CRs must be of
 extragalactic origin. 2. Around and above the spectral feature
that has been described as a bump and/or a knee, which occurs around
 $10^{15}\ev$, and starts at  $E_0 \sim (1-3)\times 10^{14}\ev$.
 The occasional bursters
 that occur inside the Galaxy--about once in a few
 hundred thousand years if burst emission is isotropic; more often, if it
 is beamed--could maintain the density of galactic cosmic rays at the
 observed level in this range. These two energy ranges might correspond
to two typical CR energy scales expected from bursters: one pertinent
to CR acceleration due to interaction of a magnetized-fireball front
with an ambient medium; the other to acceleration in the fireball itself
(e.g. shock acceleration).}
\par
Cosmic rays are thought to originate from more than one type of source,
but what exactly their sources are is still an unsettled issue
(for recent reviews see
e.g., refs \cite{pb} - \cite{axf}). Those with energies above
 $\sim 10^{19}\ev$ must be extragalactic, because
their gyroradius in galactic magnetic fields is larger than
galactic size. Those of lower-energy,
which are of galactic origin,
also seem to come from more than one type of source\cc{axf},
 as indicated by the following
observations: 1. The change in the spectral form above
 $E_0$ may witness a change in type of source, or at least in the
 acceleration mechanism.
 2. Supernova remnants (SNRs), which, on
various grounds, are thought to be adequate sources for the lower
energy CRs, seem incapable\cc{be} of
producing the higher-energy galactic CRs with
 $E>10^{14}\ev$. It may be significant that this theoretical
cutoff at $\sim 10^{14}\ev$ marks the beginning of the observed change in
both the spectrum, and the elemental composition.
 After Axford\cc{axf} we designate the former,
 sub-$E_0$ region GCRI, and the latter by GCRII.
 It is widely acknowledged (e.g. refs \cite{dr,axf})
 that the origins of both GCRII and extragalactic CRs (EGCRs)
have not been pinpointed yet. Below, we argue that \grbd\ sources (GRBSs)
may be suitable sources of both these components of CRs.
\par
Milgrom and Usov\cc{mu} have recently pointed out that the directions of
arrival of the two most energetic CRs observed to date each agree,
 within the
uncertainties, with that of an energetic \grb\ (GRB). Inspired by this
 we have proposed that ultra-high-energy CRs ($E>10^{20}\ev$)
 are accelerated in the same
cataclysmic event that produces the GRB
 (the CRs
arriving with some delay, and positional disparity
 due to wriggling of its trajectory by
 intervening magnetic fields). Waxman\cc{wa}, and Vietri\cc{viet}
 have
 made a similar proposal.
 This has, naturally, induced us to consider the
 contribution of GRBSs to the observed CR population at lower energies.
\par
The isotropy in the directions of arrival of
 \grbs, together with the distance distribution
implied by their $log\, N-log\,S$ relation, points strongly to a
GRB origin far outside the disk of the Milky Way.
This has lead to two alternative
 pictures: one in which GRBs are of cosmological
origin\cc{uc} $^-$ \cc{tm}
 and one by which their sources are distributed in a spherical,
extended corona, surrounding the Milky Way\cc{fish}
$^-$ \cc{prr}. We adopt the cosmological scenario in
the following discussion; the galactic-corona alternative
will be discussed, occasionally.
\par
 One has to consider two types of \crd\ contribution: 1. CRs at very high
energies that are hardly affected by magnetic fields in their
putative mother galaxy, and in our own, arrive here as extragalactic CRs.
2. The occasional GRBSs that occur in the
 Milky Way (still in the cosmological scenario for GRBs)
 produce a flux of high-energy CRs
passing Earth on their way out to become extragalactic CRs in other
 galaxies; these we ignore here.
They also emit CRs of low enough energies to be trapped in the Galaxy for
an extended period and contribute to its CR budget.
We first consider such trapped CRs from bursts inside the Galaxy.
\par
The observed energy density of all CRs with $E>10^{10}\ev$
 in the neighborhood of the
solar system is $w_{_{\rm CR}}\simeq
0.5\ev$ cm$^{-3}$.
 For maintaining this energy density, in spite of CR escape,
 a  refilling of the galactic volume by CRs
with a power $Q_{_{\rm CR}}^{^I}\simeq 3\times 10^{40}\es$ is
 required (see e.g. \cite{ber}).
As to GCRII, the power which is needed to maintain the observed
energy density, $w_{_{\rm CR}}^{^{II}}$ of CRs with energy above $E_0$
 is estimated at
\BE Q_{\rm CR}^{^{II}}\simeq {V_{_{\rm CR}}w_{_{\rm CR}}^{^{II}}\over
\tau_{\rm conf}(E_0)}
\simeq Q_{_{\rm CR}}^{^I}\left({E_0\over 10^{10}\ev}
\right)^{- (\gamma -2)+\eta}\;\;\es\,.\label{ii} \EE
Here $V_{_{\rm CR}}=\pi R_{\rm g}^2h_{_{\rm CR}}\simeq 10^{67}
(h_{_{\rm CR}}/h_{\rm rad})$ cm$^3$ is the volume where CRs accumulate;
$R_{\rm g}\simeq 10$ kpc is some measure of
 the radius of the Galaxy, $h_{_{\rm CR}}$
is the thickness of the CR volume; $h_{\rm rad}\simeq 750$ pc is
the thickness of the radio disk;
 $\gamma\simeq2.7$ is the spectral power of CRs
from $10^{10}\ev$ to $\sim E_0$ (ref. \cite{axf});
 so that $w_{_{\rm CR}}^{^{II}}\simeq w_{_{\rm CR}}(E_0/
10^{10}\ev)^{-(\gamma -2)}$. We have taken an energy
 dependence of the confinement time of the form

\BE \tau_{\rm conf}(E)\simeq 10^7\left({h_{_{\rm CR}}\over h_{\rm rad}}
\right)\left({E\over 10^{10}\ev}\right)^{-\eta}\;\;{\rm yr}
\label{iii} \EE
and $\eta =0.4\pm 0.2$ (refs \cite{sw} - \cite{pt}). Actually, very
little is known about the confinement time for energies outside the range
 $10^{10}-10^{12}\ev$, where it can be measured approximately from the
 abundance of radioactive secondaries among the CRs.
 In this range one result is\cc{sw}
$\eta\simeq 0.6$ . More generally, it was argued \cc{bier}\cc{pt} that
in a wide range of energies, from $\sim 10^9\ev$
to $\sim 10^{17}\ev$, the mean value of $\eta$ is about $0.3\pm 0.1$.
If we use $\eta =0.3$ and $E_0=3\times 10^{14}\ev$ we have from
 eq.(\ref{ii})

\BE Q_{_{\rm CR}}^{^{II}}\simeq 5\times 10^{38}\es. \label{iv} \EE
While the value of $h_{_{\rm CR}}$ is very uncertain, and can be anywhere
 between $\sim h_{\rm rad}$ and $\sim 2R_{\rm g}$ (e.g. ref. \cite{ber}),
our estimate of $Q_{_{\rm CR}}^{^{II}}$ does not depend on it.
\par
What fraction of such a CR luminosity is reasonable of GRBSs to supply?
To get an idea we estimate the long-time-mean luminosity in \gr s
produced by GRBSs inside the Galaxy.
 From the typical distances (as inferred from
 observed statistics), and from the fluxes of GRBs, it is estimated
(e.g. refs \cite{har}\cite{cp}\cite{wl})
 that, if the \gr s are emitted isotropically
from their sources, an average GRBS outputs $\sim 4\times
10^{51}$ ergs in \g-rays;
 and that if GRBSs occur in galaxies with rates proportional
to the galactic mass, about one GRB occurs in a galaxy like ours
once every $\sim 4\times 10^{5}$ years. If the \gr\ emission is beamed
the luminosities are smaller, in proportion, and the rate of occurrence
in a galaxy is higher, in
inverse proportion, to the beam solid angle, which might be rather small.
 By these estimates the mean power that
is released in CRs in the Galaxy from  bursts inside it is
$L_{_{\rm CR}}\simeq 3\times 10^{38}\alpha\es$, where
$\alpha$ is some typical
ratio of CR-to-\gr\ total energy output. Thus {\it if $\alpha$ is
of order unity the CR luminosity produced by the GRBs in the Galaxy
may tally with what is needed to maintain the observed CR flux
in } GCRII by eq.(\ref{iv}). In the galactic-corona-origin picture
similar arguments require $\alpha>10^4$.
\par
To achieve the observed degree of isotropy of GCRII is not much more
difficult than with other type of sources, say SNRs;
 even though GRBSs inject CRs in short,
 far-between bursts in space and time. Especially so as that the
 confinement time may be much larger than the repetition time of the
 GRBSs in the Galaxy.
\par
Consider now the possible contribution of GRBSs to the observed EGCR.
 The reckoning here is different: we now compare the integrated flux
of \gr\ bursts with that of CRs of energy $E>10^{19}\ev$.
We assume that intergalactic magnetic fields are small enough so that
such CRs are not accumulated in the IGM.
 The integral energy flux of EGCRs is
$I_{_{\rm CR}}(E>10^{19}\ev)\simeq 4\times 10^{-10}\ecms$
 above $10^{19}\ev$ and $I_{_{\rm CR}}(E>10^{20}\ev)
\simeq 0.6\times 10^{-10}\ecms$ above $10^{20}\ev$
(ref \cite{bce}).
The energy flux in GRBs is $I_\gamma
=\dot NS_0\simeq 3\times 10^{-10}\ecms$, where
$\dot N\simeq 10^3$ yr$^{-1}$ is the total observed burst rate and
$S_0\simeq 10^{-5}\ecmt$ is a typical fluence per burst,
as derived from the BATSE survey (e.g., ref \cite{har}). From
these estimates we can see that the flux of EGCRs with $E\sim 10^{19}
\ev$ may be produced by GRBSs if their luminosities in \gr s and
CRs of energies $\sim 10^{19}\ev$ are similar: $\alpha\sim
1$ at $E\sim 10^{19}\ev$.
The fact that the estimated, integrated flux of CRs above $10^{20}\ev$
is rather smaller than that above $10^{19}\ev$ does not indicate that the
spectrum is so steep at the source, as the higher-energy CRs suffer
effective degradation of energy on their way. In fact, observations
are consistent with an $E^{-2}$ energy spectrum that imparts equal
energies to equal logarithmic ranges. (See also the discussion
by Waxman\cc{wa} who reaches similar conclusions on this point).
\par
 The observed isotropy in the directions to GRBSs\cc{har}, would
 naturally explain the near isotropy of EGCRs, if these indeed are
 produced by GRBSs. There may be an added isotropization effect due to
 Larmor bending of the trajectories of the lower-energy EGCRs
 by intergalactic magnetic fields.
(The limits on such fields are so lax as to be consistent with either
an important such effect, or with hardly any.)
However, CRs with the highest energies, say $E>5\times10^{19}\ev$,
cannot, typically, come from very large distances, because their
energy is quickly degraded on travel\cc{ssb94}. Their positional
distribution will than reflect that of the nearer-by GRBSs--at these
energies the CR apparent position has a good memory of the direction of
 the source (e.g. ref. \cite{ssb94}). If GRBSs occur in galaxies, the
CR distribution will reflect any anisotropy in the distribution
of nearby galaxies (distance less than 200-300 Mpc).
 Indeed, such an anisotropy for $E>4\times10^{19}\ev$,
 correlated with the local supercluster has been found recently\cc{sbl}.
GRBSs within this distance are also far-between in time (a few per
 year). Thus we could see $E>10^{20}\ev$ CRs to come in groups
spread in time over periods of a few months to a few years
 (see ref. \cite{mu}), and concentrated in an area of the sky around the
GRBS position. This could be tested in future with the
proposed 5000 km$^2$ CR detector, the Auger observatory,
with which a few tens such CRs are expected for these strongest GRBS.
\par
What are the typical CR energies expected from GRBSs, and what
are the relative efficiencies, $\alpha$,
 of CR to \gr\ productions, which enter our deductions of the expected
 CR fluxes from \gr\ observations?
\par
There are two energy scales that might characterize the energy of cosmic
 rays accelerated in GRBSs; one is $\Gamma_0^2mc^2$, where $\Gamma_0$
is the bulk Lorentz factor of a relativistic wind, and $m$ the mass
of the particle; the other may be defined, for example by the
electric potential produced by rotating magnetic fields so that it
depends on the magnetic-field strength, and the rotation frequency.
The first scale enters as follows:
A common feature of all acceptable models of
cosmological GRBSs is that a relativistic wind is involved in
the \gr\ radiation process. The bulk Lorentz factor of such a wind
is constrained to be more than $10^2-10^3$ (refs \cite{wl},\cite{bar}).
 A very strong
magnetic field may exist in the plasma that flows away from GRBSs
\cc{npp} $^-$ \cc{us}.
It was argued by M\'esz\'aros and Rees\cc{mr}
that the X-ray and \gr\ emission
of bursts is generated far from the bursters, at distances of
$r_{\rm d}\sim 10^{15}-10^{16}$ cm, in the process of
interaction between relativistic winds with $\Gamma_0\geq 10^2$
and an external medium
(e.g., an ordinary interstellar medium, or plasma that is ejected
from the predecessor of the burster). Such a process has been
 considered numerically\cc{su}, and it was shown that the energy
that is lost by a relativistic magnetized wind because of its
interaction with an external medium is distributed in the following
way. About half of this energy is in ultra-relativistic heavy
particles, protons and nuclei  that are knocked by the wind
front. In the rest frame of the front the particles are reflected,
more-or-less elastically, from the front. Thus,
 the mean Lorentz factor of reflected heavy particles in the
frame of the burster is $\sim \Gamma_0^2$.
The other half of the wind-energy losses is distributed, more-or-less
 evenly, between low-frequency oscillations of electromagnetic
fields and both high-energy electrons and their high-frequency
(X-ray and \gr) emission.
 The total \gr\
efficiency may reach $20-30\%$.
Therefore, it is estimated that the energy released
in CRs is at least twice that in  \gr s: $\alpha \geq 2$.
 For this lower limit the estimated mean
power released in CRs in the Galaxy from local GRBSs is
$L_{_{\rm CR}} \simeq 6\times 10^{38}\es$, which is comparable
to $Q_{_{\rm CR}}^{^{II}}$ within uncertainties of their estimates.
\par
 The mean energy of CRs
knocked by the magnetized-fireball front would be the
 typical energy of the bump, $\sim 10^{15}\ev$ per nucleus,
if $\Gamma_0\sim 10^3$, which is much in keeping with what is expected;
but, $\Gamma_0$ may be as high as $10^{5}$ (ref. \cite{mr}) giving
CR energies of up to $\sim 10^{19}\ev$ per nucleus.
 Hence, the occasional GRBSs that occur in the
Milky Way might explain both the energetics, and the typical
energy of GCRII.
 Our limited knowledge of the processes involved does not suffice do
determine the expected spectrum of CRs.
\par
The second energy scale enters because
CRs may be generated by mechanisms other than through knocking by the
relativistic interface
between the outflowing gas and an external medium, as described above.
 For example, by one class of cosmological-origin
 models, GRBs are produced in rotating disk-like objects resulting from
 mergers of a neutron-star binaries (e.g. ref. \cite{pir94}), or by
 neutron stars formed by accretion-induced collapse\cc{us}. The
magnetic field, $B_{_{\rm S}}$, at the surface of such objects
may be as high as $\sim 10^{16}$ G (refs \cite{npp},
\cite{us}), and their angular velocity $\Omega \sim 10^4$ s$^{-1}$.
 The potential
difference between the surface of such an object, at radius
 $R\sim10^6$cm, and infinity is\cc{rs75}
\BE \Delta \varphi_{\rm max}={\Omega^2B_{_{\rm S}}R^3\over 2c^2}
\sim  10^{23}\left({\Omega\over 10^4\,{\rm s}^{-1}}
\right)^2\left({B_{_{\rm S}}\over 10^{16}\,{\rm G}}\right)
\left({R\over 10^6\,{\rm cm}}\right)^3\;\;{\rm V}\,. \label{v} \EE
Charged particles that flow away from the surface
 may be accelerated, in principle, up to
the energy $E_{\rm max}\simeq e\Delta \varphi_{\rm max}$,
enough to explain EGCRs of all observed energies.
Alternatively, particles may be accelerated by relativistic shocks
that may be formed in an unsteady relativistic wind\cc{rm94}
 up to similar
 energies (see e.g. refs \cite{mu,wa,viet}).
\par
 For very-high-energy CRs (termed EGCR) to originate
in coronal GRBSs requires a CR to \gr\ ratio $\alpha\sim 1$, but
existing physical GRBS models (e.g. ref. \cite{prr})
 are incapable of producing CRs with such high
energies.

%\clearpage***
\end{document}